\documentclass[epj]{svjourmod}

\usepackage{amsmath,amssymb,amsfonts}
\usepackage{graphicx,color}

\newcommand{\beq}{\begin{equation}}
\newcommand{\eeq}{\end{equation}}
\newcommand{\D}{\displaystyle}
\newcommand{\gev}{\, \text{GeV}}

\newcommand{\piprime}{\Pi^{\,\prime}}
\newcommand{\disc}{{\rm disc}\,}
\renewcommand{\Re}{{\rm Re}\,}
\renewcommand{\Im}{{\rm Im}\,}
\renewcommand{\arg}{{\rm arg}\,}
\newcommand{\nnnl}{\nonumber\\}
\newcommand{\diff}{\text{d}}

\begin{document}

\title{Constraints on the \boldmath{$\omega\pi$} form factor from analyticity and unitarity}
\titlerunning{Constraints on the $\omega\pi$ form factor\ldots}

\author{B. Ananthanarayan\inst{1} \and I. Caprini\inst{2} \and B. Kubis\inst{3}} 
\institute{
   Centre for High Energy Physics,
   Indian Institute of 
   Science, Bangalore 560 012, India 
\and
   Horia Hulubei National Institute for Physics and Nuclear Engineering,
   P.O.B. MG-6, 077125 Magurele, Romania
\and
   Helmholtz-Institut f\"ur Strahlen- und
   Kernphysik (Theorie) and Bethe Center for Theoretical Physics,
   Universit\"at Bonn, D-53115 Bonn, Germany
}

\authorrunning{B.\ Ananthanarayan et al.} 

\date{
}

\abstract{Motivated by the discrepancies noted recently between the theoretical calculations  of the electromagnetic $\omega\pi$ form factor and certain experimental data,  we investigate this form factor 
using analyticity and unitarity in a framework known as the method of unitarity bounds.
We use a QCD correlator computed on the spacelike axis by operator product expansion and
perturbative QCD as input, and exploit unitarity and the positivity of its spectral 
function, including the two-pion contribution 
that can be reliably calculated using high-precision data on the pion form factor.
From this information, we derive upper and lower bounds on the modulus of the $\omega\pi$ form factor 
in the elastic region. The results provide a significant check on those
obtained with standard dispersion relations, confirming the existence of a disagreement with experimental 
data in the region around $0.6\gev$.
}

\maketitle
\section{Introduction}

Recent years have seen strongly increased interest in the transition form factors of light mesons~\cite{Czerwinski:2012ry}.
One of the main reasons is the fact that pseudoscalar ($\pi^0, \eta, \eta'$) pole terms constitute some of the
most important contributions to hadronic light-by-light scattering, which is soon to become the biggest stumbling block
in a more accurate theoretical determination of the standard model prediction for the muon's anomalous magnetic moment;
see~\cite{Jegerlehner:2009ry} and references therein.  The strength of these pole terms is determined by the 
singly- and doubly-virtual form factors for $\pi^0,\,\eta,\,\eta' \to \gamma^*\gamma^{(*)}$.
However, for the virtuality of one of the photons fixed to the mass of one of the light isoscalar vector resonances
$\omega$ or $\phi$, these form factors are intimately linked to vector-meson transition form factors
that can be measured in decays such as $\omega\to\pi^0\ell^+\ell^-$ or $\phi\to\eta\ell^+\ell^-$.
In all likelihood, these vector-meson conversion decays present one of the few opportunities to measure the 
doubly-virtual $\pi^0$ or $\eta$ form factors with very good precision.
Furthermore, they present an essential ingredient to a more advanced theoretical understanding of hadronic
light-by-light scattering~\cite{roadmap,pi0TFF}.

\begin{sloppypar}
Recent dispersive treatments~\cite{Schneider:2012ez,Danilkin:2014cra} of the $\omega\pi$ electromagnetic form factor $f_{\omega\pi}(t)$  are in disagreement with experimental data in the region around $0.6\gev$~\cite{LeptonG,NA60,NA60new}, which show strong deviations from even approximate vector-meson-dominance behavior~\cite{Landsberg}. The main ingredient in the dispersion relation  is unitarity, which allows one to express the discontinuity of the form factor in terms of the P~partial wave of the process $\pi\pi\to\omega\pi$~\cite{Niecknig:2012sj,Danilkin:2014cra} and the pion electromagnetic form factor, quantities determined with precision. Strictly speaking, this relation is valid only in the elastic region, $4 m_\pi^2\leq t<16m_\pi^2$. Due to the strong phase-space (and chiral) suppression of multiparticle intermediate states, the elastic regime approximately extends up to $\omega\pi$ threshold in the P~wave,  $4 m_\pi^2\leq t<t_+ = (m_\omega+m_\pi)^2$, above which the inelasticity in pion--pion scattering is assumed to be dominated by the $\omega\pi$ intermediate state~\cite{Niecknig:2012sj}.  However, due to the lack of information on the discontinuity in the inelastic region, elastic unitarity is assumed to be valid also at higher energies in the evaluation of  the dispersion integral. This assumption may affect the precision of the theoretical treatment. Having in view the disagreement with some experimental data on $|f_{\omega\pi}(t)|$ below $t_+$, it is of interest to investigate the form factor in a more model-independent framework, which avoids this assumption.

In the present paper we exploit alternative information on the form factor above  $t_+$. We use a method proposed originally by Okubo~\cite{Okubo1,Okubo2} (before the advent of QCD), which leads to bounds on form factors by exploiting the positivity of the spectral function of a suitable current--current correlator. This technique, known as \textit{method of unitarity bounds}, has been resuscitated in the QCD era and was applied to a variety of form factors of heavy and light mesons~\cite{BoMaRa,deRafael:1993ib,Lellouch:1995yv,Boyd:1995cf,CaMa,CaLeNe,IC,Caprini:2010ye} (for a review and more references see~\cite{Abbas:2010EPJA,AC}).  In the present study, we use  a dispersion relation for the polarization function of two isovector vector currents, calculated by operator product expansion (OPE) in the Euclidean region, and exploit unitarity for the spectral function. Including $\pi\pi$ and $\omega\pi$ states in the unitarity sum, we  derive an upper bound for an integral on the modulus squared of the $\omega\pi$ form factor along the cut from $t_+$ to infinity. From this condition and the known discontinuity in the unphysical region  $4 m_\pi^2\leq t< t_+$, we derive bounds on the form factor in the latter region and compare them with the experimental data and the results of the standard dispersion relations. 

In this work, we have adapted the original method of unitarity bounds, making it suitable for the information available on the $\omega\pi$ form factor, namely its discontinuity across the cut below the inelastic threshold. To solve the corresponding optimization problem we use analytic techniques already applied to scattering amplitudes~\cite{Caprini:1981,Caprini:1982nr}. We also had to take into account the fact that, unlike most form factors studied up to now, the $\omega\pi$ form factor is not a real analytic function.\footnote{A function $F(t)$ analytic in the  $t$-plane cut for $t\ge 4 m_\pi^2$ is of real type if it satisfies the condition $F(t^*)= (F(t))^*$. In particular, this implies that the function is real on the real axis for $t<4 m_\pi^2 $, while its discontinuity across the cut can be written as $\disc F(t)\equiv  F(t+i\epsilon)- F(t-i\epsilon)=2 i \, \Im F(t+i\epsilon)$.} We have therefore made the proper generalization of the formalism to the case of analytic functions that are not of real type. To our knowledge such
a formulation appears here for the first time and represents an important generalization of methods existing in the literature. The above modifications may have applications in the study of other form factors as well. 
\end{sloppypar}

In Sect.~\ref{sec:dr} we briefly summarize the standard dispersive treatment of the $\omega\pi$ form factor. In Sect.~\ref{sec:unit} we review the formalism of unitarity bounds, which leads to the integral constraint on the modulus squared of the form factor along the cut. In Sect.~\ref{sec:formalism} we solve the optimization problem and derive upper and lower bounds on $|f_{\omega\pi}(t)|$ for $t$ below $t_+$. Section~\ref{sec:res} contains the numerical results of our work and 
Sect.~\ref{sec:disc} our conclusions.

\section{Standard dispersive treatment of \boldmath{$f_{\omega\pi}(t)$}}\label{sec:dr}
We use the definition from~\cite{Schneider:2012ez}, where the form factor $f_{\omega\pi}(t)$ is defined from the matrix element
\begin{equation}\label{eq:fdef}
\langle \omega(p_a,\lambda)\pi(p_b)| j_\mu(0) | 0 \rangle = i
\epsilon_{\mu\tau\rho\sigma}\epsilon^{\tau *}(p_a, \lambda) p_b^\rho q^\sigma f_{\omega\pi}(t),
\end{equation}
where $j_\mu$ is the isovector part of the electromagnetic current, $\lambda$ denotes the $\omega$ polarization, and we defined
 $q=p_a+p_b$ and $t=q^2$.
The form factor $f_{\omega\pi}(t)$  has dimension ${\rm GeV}^{-1}$. The definition adopted in~\cite{Koepp:1974da} contains a factor $1/m_{\omega}$ in the right-hand side of~\eqref{eq:fdef} and 
 the corresponding form factor is dimensionless.  In the present paper
we will work with the dimensionful version of the form factor.

\begin{sloppypar}
Unitarity implies that $f_{\omega\pi}(t)$ has a cut along the real axis for $t \ge 4 m_\pi^2$.
Using the conventions of~\cite{Schneider:2012ez}, the discontinuity of $f_{\omega\pi}(t)$ across the cut in the elastic approximation is given by
\begin{equation}\label{eq:disc}
 \disc{f_{\omega\pi}(t)}=\frac{i\,q_{\pi\pi}^3(t)}{6\pi \sqrt{t}}F_\pi^{*}(t)f_1(t)\,\theta\big(t-4 m_\pi^2\big) ,
\end{equation}
where $q_{\pi\pi}(t)=\sqrt{t/4 - m_\pi^2}$, $F_\pi(t)$ is the pion electromagnetic form factor, and $f_1(t)$ the P~partial-wave amplitude of the scattering process
\beq
\pi^+(q_1) \, \pi^-(q_2)\to \omega(p_a,\lambda) \, \pi^0(p_b).
\eeq
In~\cite{Koepp:1974da} the partial wave $f_1(t)$ was calculated by the $N/D$ method, with the left-hand cut approximated by $\rho$-exchange. In this model, the phase of  $f_1(t)$ exactly compensates the phase of $F^*_\pi(t)$ in the discontinuity~\eqref{eq:disc},  and as a consequence the form factor $f_{\omega\pi}(t)$ is a real analytic function. 
However, as discussed in~\cite{Schneider:2012ez}, once rescattering effects are taken into account, the phase of the partial wave $f_1(t)$ no longer coincides with the $\pi\pi$ P-wave phase shift. More precisely, in the projection onto the P~partial wave, the kinematical variables reach regions where the decay 
\beq
\omega(p_a,\lambda)\to \pi^+(q_1) \, \pi^-(q_2) \, \pi^0(-p_b)
\eeq
is allowed and rescattering between the final pions including three-pion cuts is possible. As a consequence, the discontinuity~\eqref{eq:disc} is not purely imaginary and the $\omega\pi$ form factor is not a real analytic function.

In~\cite{Schneider:2012ez} the pion vector form factor has been reconstructed from an Omn\`es representation~\cite{Omnes} using pion--pion phase shifts~\cite{Madrid,CCL} as input. The partial wave $f_1$ was calculated in~\cite{Niecknig:2012sj} using Khuri--Treiman techniques~\cite{KhuriTreiman}.\footnote{Reference~\cite{Danilkin:2014cra} is a variant of the calculations~\cite{Niecknig:2012sj,Schneider:2012ez}, whose differences to those studies are immaterial for the present investigation.}
Besides not fulfilling a straightforward variant of Watson's final-state phase relation~\cite{Watson}, it also has the peculiarity of showing a singular behavior at the pseudothreshold $t_-=(m_\omega-m_\pi)^2$, which can be understood perturbatively from the analytic structure of certain non-trivial two-loop Feynman diagrams (see e.g.~\cite{cusp}).  Note that this does not imply any singular behavior of $f_{\omega\pi}(t)$ near $t_-$: the form factor remains regular on the upper rim of the cut. 
\end{sloppypar}

The expression~\eqref{eq:disc} is valid in the region $4 m_\pi^2\leq t< (m_\omega+m_\pi)^2$, since above the $\omega\pi$ threshold other intermediate states contribute in the unitarity sum. By neglecting these contributions, the form factor was obtained from a once-subtracted dispersion relation~\cite{Schneider:2012ez,Koepp:1974da} 
\beq\label{eq:DR}
f_{\omega\pi}(t)= f_{\omega\pi}(0) + \frac{t}{2\pi i}\int_{4 m_\pi^2}^\infty \frac{ \disc{f_{\omega\pi}(t')}}{t'(t-t')} \diff t',
\eeq
where $|f_{\omega\pi}(0)|$ is known experimentally from the $\omega\to\pi^0\gamma$ decay rate. The updated value is~\cite{PDG}
\beq\label{eq:f0}
|f_{\omega\pi}(0)|=(2.30 \pm 0.04) \gev^{-1}.
\eeq

\begin{sloppypar}
 The recent analysis performed in~\cite{Schneider:2012ez}, based on the dispersion relation~\eqref{eq:DR}, leads to results that are inconsistent with some experimental data around $\sqrt{t} \approx 0.6 \gev$~\cite{LeptonG,NA60,NA60new}.  As discussed above, one questionable point of the theoretical analysis is the extension of the elastic unitarity relation~\eqref{eq:disc} above the (effective) threshold $t_+=(m_\omega+m_\pi)^2$ of multiparticle production. In the next section we shall show how some information on the modulus of the form factor $f_{\omega\pi}(t)$ for $t>t_+$ can be derived from independent sources.

\section{Consequences of perturbative QCD, analyticity, and unitarity}\label{sec:unit}

\begin{sloppypar}
Using standard techniques~\cite{Okubo1,Okubo2,BoMaRa,deRafael:1993ib,Lellouch:1995yv,Boyd:1995cf,CaMa,CaLeNe,IC,Caprini:2010ye,Abbas:2010EPJA,AC}, we start with the QCD
vacuum polarization tensor 
\begin{align}\label{eq:correl}
\Pi^{\mu\nu}(q) &= \int \diff x\, e^{iqx} \langle 0 | T[j^\mu(x) j^\nu(0)]|0\rangle \nnnl
& =\big(q^\mu q^\nu -g^{\mu\nu} q^2\big)\Pi(t), \quad\quad t=q^2,
\end{align}
where $j_\mu$ is the isovector part of the electromagnetic current. The first derivative $\piprime(t)$ of the QCD vacuum polarization amplitude 
$\Pi(t)$ satisfies the  dispersion relation
\beq
\piprime(t) = \D\frac{1}{\pi} \int_0^{\infty} \frac{\Im \Pi(t' + i\epsilon)}{(t' - t)^2} \,\diff t',
	\label{eqn:disp1}
\eeq
with the spectral function given by the unitarity relation
\begin{align}
& \big(q^\mu q^\nu -g^{\mu\nu} q^2\big)\Im \Pi(t+i \epsilon)
= \nnnl
&
\frac{1}{2} \sum_\Gamma \int \diff\rho_\Gamma (2 \pi)^4 \delta^{(4)}(q-p_\Gamma)
\langle 0 | j^\mu(0) | \Gamma \rangle \langle \Gamma | j^\nu(0)^\dag |
0 \rangle. \label{eq:CorrUnit}
\end{align}
Keeping the $\pi\pi$ and $\omega\pi$ intermediate states explicitly (which should dominate the isovector
spectral function at low energies), carrying out 
the two-body phase space integrals and using the positivity of the spectral function, we obtain the inequality
\begin{align}\label{eq:posit}
\piprime(t)  &\ge  \int_{4 m_\pi^2}^\infty w_\pi(t',t) |F_\pi(t')|^2 \diff t'   \nnnl & + \int_{t_+}^\infty
 w_{\omega\pi}(t',t)  |f_{\omega\pi}(t')|^2 \diff t',
\end{align}
where 
\begin{align}\label{eq:w}
w_\pi(t',t)&= \frac{1}{48 \pi^2} 
\frac{ 1 }{(t'-t)^2} \left(1-\frac{4 m_\pi^2}{t'}\right)^{3/2}, \\
w_{\omega\pi}(t',t) &= \frac{1}{192 \pi^2} 
\frac{t'} {(t'-t)^2} \left(1-\frac{t_-}{t'}\right)^{3/2} 
\left(1-\frac{t_+}{t'}\right)^{3/2}.\nonumber
\end{align}
It is convenient to write~\eqref{eq:posit} as an integral constraint on the modulus of the $\omega\pi$ form factor as  
\begin{equation}\label{eq:L2}
  \int_{t_+}^\infty
 w_{\omega\pi}(t',t)  |f_{\omega\pi}(t')|^2 \diff t' \leq I(t) ,
\end{equation}
where
\beq\label{eq:I}
I(t)= \piprime(t) - \int_{4 m_\pi^2}^\infty w_\pi (t',t)|F_\pi(t')|^2 \diff t'.
\eeq
 This quantity can be evaluated for spacelike values $t\equiv -Q^2<0$ using OPE and perturbative QCD for the correlator $\piprime(t)$, and the rich information available on the modulus of the pion form factor. The value of $Q^2$ should be taken large enough such as to ensure the validity of the OPE, and in the same time lead to sufficiently strong bounds. As discussed in~\cite{IC}, a reasonable choice satisfying these requirements is $Q^2=2\gev^2$. We used perturbative QCD to four loops (see~\cite{QCD} and references therein):
\begin{align}\label{eq:qcd}
\piprime_{\text{pert}}(-Q^2)=\frac{1}{8 \pi^2 Q^2} \Big( &1  + 0.318 \alpha_s + 0.166 \alpha_s^2 \nnnl
 &+  0.205 \alpha_s^3+ 0.504 \alpha_s^4\Big), 
\end{align}
where $\alpha_s$ is the strong coupling at $Q^2=2\gev^2$.  Using as input the value $\alpha_s(m_\tau^2)=0.320 \pm 0.020$, which covers most of the recent determinations from hadronic $\tau$ decays (see for instance~\cite{Beneke:2008,CaFi2011,Boito_update,Abbas:2013,Boito_2014}), and the coupling's running we obtain $\alpha_s=0.357 \pm 0.025$. This yields for $\piprime_{\text{pert}}$ the central value $0.0073 \gev^{-2}$ with an error of about 1.3\%.  We also checked that a higher-order term added in~\eqref{eq:qcd}, taken as $0.925\alpha_s^5$ according to~\cite{Beneke:2008}, changes $\piprime_{\text{pert}}$ by about 1.2\%.  The power corrections in the OPE, obtained from~\cite{Braaten:1992}, bring a small contribution. For instance, using for the gluon condensate the standard value $\langle\alpha_s G^2\rangle/\pi = 0.012\gev^4$ given in~\cite{SVZ1,SVZ2}, we obtain for its contribution the value $0.0001 \gev^{-2}$. This leads to $\piprime(-2\gev^2)=(0.0074 \pm 0.0001) \gev^{-2}$, where the uncertainty includes quadratically the effects of the $\alpha_s$ uncertainty and the truncation error.
\end{sloppypar}

The integral involving the pion electromagnetic form factor can be
calculated using in the low-energy region BaBar experimental data~\cite{BaBar} and the bounds on $|F_\pi(t)|$  derived in~\cite{Ananthanarayan:2012tt}, along with data obtained by BaBar up to $3\gev$~\cite{BaBar} and a smooth transition to the $1/t$ decrease predicted by QCD (for details see~\cite{Ananthanarayan:2012tt,Ananthanarayan:2012}).  This gives for the integral appearing in~\eqref{eq:I} the value $(0.0033\pm 0.0001) \gev^{-2}$, which leads to 
\beq\label{eq:Ierror}
I\equiv I(-2\gev^2) =( 0.0041 \pm 0.0002) \gev^{-2}.
\eeq
From the inequality~\eqref{eq:L2} and the discontinuity~\eqref{eq:disc} adopted in the elastic region of validity $t<t_+$,  we shall obtain bounds
on the form factor at points below $t_+$. The mathematical technique will be explained in the next section. 
\end{sloppypar}

\section{Bounds on \boldmath{$|f_{\omega\pi}(t)|$} below \boldmath{$t_+$}}\label{sec:formalism}

In this section, we sequentially formulate the extremal problem, present
its solution, and also discuss the special case of the optimal
solution for the case of a real analytic form factor.

\subsection{Formulation of an extremal problem}
 In order to cast the problem into a canonical form,  the first step is to 
map the  $t$ plane cut along $t\ge t_+$  onto the unit disk $|z|<1$ in the  $z\equiv \tilde z(t)$ plane, by using a suitable
conformal mapping. In particular we shall use the mapping 
\beq\label{eq:z}
\tilde z(t)=\frac{1-\sqrt{1-t/t_+}}{1+\sqrt{1-t/t_+}},
\eeq 
such that the $\tilde z(0)=0$. In the $z$-plane the elastic region $4 m_\pi^2\le t<t_+$ becomes the segment $x_\pi\le x<1$ of the real axis, where $x_\pi=\tilde z(4 m_\pi^2)$,  and the upper (lower) edges of the cut $t>t_+$ become the upper (lower) semicircles. 

\begin{sloppypar}
Further, we shall construct an outer function $C(z)$, {\em i.e.} a function analytic and without zeros in $|z|<1$, its modulus  on $|z|=1$ being equal to $\sqrt{w_{\omega\pi}(\tilde t(z), -Q^2) |d\tilde t(z)/dz|}$, where $\tilde t(z)$ is the inverse of~\eqref{eq:z}. The construction of the outer functions is explained in~\cite{Duren} (see also the review~\cite{Abbas:2010EPJA}).  Using the expression~\eqref{eq:w} of $w_{\omega\pi}(t, -Q^2)$, we obtain for $C(z)$ the exact analytic expression
\begin{align}\label{eq:C}
C(z)& =  \frac{(1-z)^{2}
(1+z)^{-1/2}}{ 16 \sqrt{6} \pi} \nonumber\\
&\times  \frac{(1+\tilde{z}(-Q^2))^2  
 (1-z\, \tilde{z}(t_-))^{3/2}}{ 
 (1-z \,\tilde{z}(-Q^2))^2
(1+\tilde{z}(t_-))^{3/2}} \,.
\end{align}
Then the inequality~\eqref{eq:L2},  written  in terms of the new function $h(z)$ defined as
\beq\label{eq:h}
 h(z) = C(z)\,f_{\omega\pi}(\tilde t(z)),
\eeq
 becomes\footnote{The  outer function ensures that~\eqref{eq:L2h} is fully equivalent with the original inequality~\eqref{eq:L2}~\cite{Duren}.} 
\beq\label{eq:L2h}
 \D\frac{1}{2 \pi} \int_0^{2 \pi} \diff\theta |h(e^{i\theta})|^2 \leq I,
\eeq
where $\theta=\arg z$.
\end{sloppypar}
Since $C(z)$ is real analytic in $|z|<1$, $C(x)$ is real for $ x_\pi \le x<1$, and from the definition~\eqref{eq:h} it follows that we can write
\beq\label{eq:disch}
\disc{h(x)} \equiv \Delta (x)= C(x)\, \disc{f_{\omega\pi}(\tilde t(x))},
\eeq
where the discontinuity of the form factor is obtained from~\eqref{eq:disc}. The function $h(z)$ can be  expressed in terms of its discontinuity as 
\beq\label{eq:defg}
h(z)= \frac{1}{2\pi i}\int_{x_\pi}^1 \frac{\Delta (x)}{x-z} \diff x + g(z),
\eeq
where
the function $g(z)$ is analytic in $|z|<1$, as its discontinuity vanishes:
\beq\label{eq:discg}
\disc{g(x)}=0, \quad -1<x<1.
\eeq
Since we consider in general form factors that are not real analytic, the function $g(z)$ is analytic, but its values on the real axis may be complex. 

\begin{sloppypar}
We now express the available information on the form factor  as a number of constraints on the function  $g$.  By inserting~\eqref{eq:defg} in~\eqref{eq:L2h} we obtain the condition
\beq\label{eq:L2g}
\frac{1}{2 \pi}\int_0^{2 \pi} \diff\theta \left|\frac{1}{2\pi i}\int_{x_\pi}^1 \frac{\Delta (x)}{x- e^{i\theta}} \diff x + g(e^{i\theta})\right|^2\leq I, 
\eeq
while using~\eqref{eq:f0} we see that $g(0)$ has the value
\beq\label{eq:g0}
g(0)= \frac{f_{\omega\pi}(0)}{C(0)}-\frac{1}{2\pi i}\int_{x_\pi}^1 \frac{\Delta (x)}{x}\diff x. 
\eeq
The problem is to find 
the maximal allowed range of $|g(z_1)|$ at an arbitrary given point $z_1=\tilde z(t_1)$ in the interval $(x_\pi, 1)$, for functions $g(z)$ analytic in $|z|<1$ and subject both to the boundary condition~\eqref{eq:L2g} and the additional constraint~\eqref{eq:g0}.
\end{sloppypar}

Let us denote
\beq\label{eq:gz1}
g(z_1)= \xi,
\eeq
where $\xi$ is an unknown parameter. A simple reasoning  (see for instance~\cite{Caprini:1982nr}) shows that the allowed  range of  $\xi$ is described by the inequality
\beq\label{eq:range}
\mu_2^2(\xi) \leq I,
\eeq 
where $\mu_2^2(\xi)$ is the solution of the minimization problem
\beq\label{eq:mu2}
\mu_2^2(\xi) =\min_{g\in {\cal G}_\xi} \frac{1}{2 \pi}\int_0^{2 \pi} \diff\theta \left|\frac{1}{2\pi i}\int_{x_\pi}^1 \frac{\Delta (x)}{x- e^{i\theta}} \diff x + g(e^{i\theta})\right|^2
\eeq
on the class  ${\cal G}_\xi$ of functions analytic in $|z|<1$, which satisfy the constraint~\eqref{eq:g0} and the additional condition~\eqref{eq:gz1} for a given $\xi$.

\subsection{Solution of the extremal problem}

\begin{sloppypar}
We solve the constrained minimum norm problem~\eqref{eq:mu2} by the technique of Lagrange multipliers. We use the fact that the  $L^2$ norm squared of a complex function  $F(\theta)$ given on the boundary of the unit disk is expressed as
\beq
\frac{1}{2 \pi} \int_0^{2 \pi} |F(\theta)|^2 \diff\theta=\sum_{n=-\infty}^\infty |F_n|^2
\eeq
in terms of its Fourier coefficients 
\beq
F_n= \frac{1}{2 \pi}\int_0^{2 \pi} F(\theta) e^{-i n \theta} \diff\theta, \quad n \in \mathbb{Z}.
\eeq
Therefore, we write the Lagrangian of the constrained minimization problem~\eqref{eq:mu2} as
\beq\label{eq:L}
{\cal L}=\sum_{n=1}^\infty |c_n|^2+ \sum_{n=0}^\infty |g_n|^2+\lambda\bigg(\sum_{n=0}^\infty g_n z_1^n-\xi\bigg),
\eeq
where $c_n$ are the negative-frequency coefficients of the function, which can be written by applying Cauchy's theorem as
\beq\label{eq:cn}
c_n=-\frac{1}{2 \pi i} \int_{x_\pi}^1 \Delta (x) x^{n-1} \diff x, \quad n\geq 1,
\eeq
 and $g_n$ are the positive-frequency coefficients, defined through the Taylor expansion of the analytic function $g(z)$:
\beq\label{eq:gn}
g(z)=\sum_{n=0}^\infty g_n z^n. 
\eeq
We must find the minimum of the Lagrangian~\eqref{eq:L}  with respect to the complex parameters $g_n$, which are free, except $g_0=g(0)$ which is known from~\eqref{eq:g0}.
The minimum conditions 
\beq\label{eq:gnmin}
\frac{\partial {\cal L}}{\partial g_n^*}=0, \quad n\ge 1,
\eeq
have the solutions
\beq\label{eq:gnsol}
g_n= -\lambda z_1^n,\quad n\ge 1,
\eeq
which, introduced in the constraint~\eqref{eq:gz1}, give the Lagrange multiplier
\beq\label{eq:lambda}
\lambda=-\frac{1-z_1^2}{z_1^2}\,(\xi -g(0))
\eeq
and the optimal coefficients
\beq\label{eq:gnsol1}
g_n= (1-z_1^2)(\xi_1 -g(0)) z_1^{n-2},\quad n\ge 1.
\eeq
By inserting these coefficients in~\eqref{eq:L} and performing the summation of the first term, we find the minimum norm
\begin{align}\label{eq:min}
\mu_2^2(\xi) &= \frac{1}{4 \pi^2} \int_{x_\pi}^1\int_{x_\pi}^1\frac{\Delta(x) \Delta^*(y)}{1-xy} \diff x\,\diff y \nonumber\\
 &+ |g(0)|^2+ \frac{1-z_1^2}{z_1^2}\,|\xi -g(0)|^2.
\end{align}
Then the inequality~\eqref{eq:range} can be written in terms of the unknown quantity $\xi$ as
\beq\label{eq:range1}
|\xi -g(0)|\leq \frac{z_1}{\sqrt{1-z_1^2}}\,I',
\eeq
where
\beq\label{eq:Iprime}
I'=\left[I-\frac{1}{4 \pi^2} \int_{x_\pi}^1\int_{x_\pi}^1\frac{\Delta(x) \Delta^*(y)}{1-xy} \diff x\,\diff y - |g(0)|^2\right]^{1/2}.
\eeq
By using the definition~\eqref{eq:gz1} of $\xi$  and the relations~\eqref{eq:h} and~\eqref{eq:defg}, we write the inequality~\eqref{eq:range1} as
\beq\label{eq:range2} 
\left|f_{\omega\pi}(t_1) C(z_1)- \frac{1}{2\pi i}\int_{x_\pi}^1 \frac{\Delta (x)}{x-z_1} \diff x -g(0)\right|\leq \frac{z_1}{\sqrt{1-z_1^2}}\,I',
\eeq
which leads to  upper and lower bounds on $|f_{\omega\pi}(t_1)|$:
\begin{align}\label{eq:rangef}
 |f_{\omega\pi}(t_1)|&\leq \frac{\left|g(0)+\frac{1}{2\pi i}\int_{x_\pi}^1 \frac{\Delta (x)}{x-z_1} \diff x\right|+\frac{z_1 I'}{\sqrt{1-z_1^2}}}{C(z_1)}\,,\nonumber\\
|f_{\omega\pi}(t_1)|&\geq \frac{\left|g(0)+\frac{1}{2\pi i}\int_{x_\pi}^1 \frac{\Delta (x)}{x-z_1} \diff x\right|-\frac{z_1 I'}{\sqrt{1-z_1^2}}}{C(z_1)}.
\end{align}
We have taken into account  the fact that $C(z_1)$  calculated from~\eqref{eq:C} is positive for real $z_1$.  Using the value of $g(0)$ from~\eqref{eq:g0} and $I'$ from~\eqref{eq:Iprime}, with $\Delta(x)$ defined in~\eqref{eq:disch}, we have all the ingredients to evaluate~\eqref{eq:rangef} numerically at an arbitrary point $t_1<t_+$.
\end{sloppypar}

\subsection{Optimal solution for real analytic functions}
It is of interest to consider in particular the case when the form factor is a real analytic function, {\em i.e.} it  satisfies $f_{\omega\pi}(t^*) =(f_{\omega\pi}(t))^*$. A similar, more general, optimization problem for real analytic functions was  solved previously in~\cite{Caprini:1981,Caprini:1982nr} with a slightly different method. Below we shall use  the  method of Lagrange multipliers applied above, adapting it to real analytic functions. In this case the discontinuity of the form factor across the cut is equal to $ 2 i \,\Im f_{\omega\pi} (x+i\epsilon)$, and from~\eqref{eq:disch} we obtain the discontinuity  $\Delta(x)$ of $h(x)$ for $x\in (x_\pi, 1)$ as 
\beq\label{eq:Delta}
\Delta(x)=2 i \,\sigma(x),
\eeq
 where 
\beq\label{eq:sigma}
\sigma(x)= \Im h(x+i\epsilon)= C(x)\ \Im f_{\omega\pi}(\tilde t(x)+i\epsilon).
\eeq
 Moreover, from~\eqref{eq:defg} it follows that the parameter $\xi$ defined in~\eqref{eq:gz1} is real and given by
\beq\label{eq:xir}
\xi=\Re h(z_1)- \frac{P}{\pi} \int_{x_\pi}^1 \frac{\sigma (x)}{x-z_1} \diff x,
\eeq
where $P$ denotes the Cauchy principal value.  Therefore, from~\eqref{eq:range1} we obtain a quadratic inequality with real coefficients for the parameter $\Re h(z_1)$
\beq\label{eq:quadr}
\left(\Re h(z_1)- \frac{P}{\pi} \int_{x_\pi}^1 \frac{\sigma (x)}{x-z_1} \diff x-g(0)\right)^2\leq \frac{z_1^2}{1-z_1^2} I'^2,
\eeq
where $g(0)$ is now real and $I'$ can be written as
\beq\label{eq:Iprime1}
I'=\left[I-\frac{1}{\pi^2} \int_{x_\pi}^1\int_{x_\pi}^1\frac{\sigma(x) \sigma(y)}{1-xy} \diff x\,\diff y - g(0)^2\right]^{1/2}.
\eeq
 From~\eqref{eq:quadr} we obtain exact upper and lower bounds on the real part of the form factor:
\begin{align}\label{eq:boundreal}
\Re f_{\omega\pi}(t_1)  &\leq \frac{g(0)+\frac{P}{\pi} \int_{x_\pi}^1 \frac{\sigma (x)}{x-z_1} \diff x +\frac{z_1 I'}{\sqrt{1-z_1^2}}}{C(z_1)} \,,\nonumber\\
\Re f_{\omega\pi}(t_1)  &\geq \frac{g(0)+\frac{P}{\pi} \int_{x_\pi}^1 \frac{\sigma (x)}{x-z_1} \diff x -\frac{z_1 I'}{\sqrt{1-z_1^2}}}{C(z_1)}\,.
\end{align}
We have checked that these bounds follow as particular cases from the more general expressions given  in~\cite{Caprini:1981,Caprini:1982nr}.
By combining the bounds~\eqref{eq:boundreal} with the known value of the imaginary part, we derive bounds on the modulus of the form factor.  These bounds are optimal, unlike the bounds given in~\eqref{eq:rangef}, where the treatment of the modulus in the last step of the derivation amounts to a loss of optimality. 

\begin{sloppypar}
 We end this section with a remark that might be useful for improving the bounds. From general arguments~\cite{Abbas:2010EPJA} and the expressions given above, it follows that the bounds  depend monotonically on the value of $I$ in the $L^2$-norm constraint~\eqref{eq:L2}: smaller values of  $I$ lead to narrower allowed intervals for  $|f_{\omega\pi}(t)|$ at $t<t_+$.
 Therefore, the bounds can be made tighter in principle by taking into account more intermediate states, besides the $\pi\pi$ pairs, in the unitarity relation~\eqref{eq:CorrUnit} for the QCD correlator. Many of these are (re)measured exclusively with excellent precision in the ongoing quest to improve on the hadronic vacuum polarization contribution to the muon's anomalous magnetic moment (see e.g.~\cite{Benayoun:2014tra}). The positive contribution of these states can be subtracted from the QCD value of the correlator as in~\eqref{eq:I}, reducing the value of $I$.  To establish the practical effect on the bounds requires further investigations. 
\end{sloppypar}

\section{Results}\label{sec:res}

In the calculation of the bounds
we have employed the discontinuity of the $\omega\pi$ form factor in the range $(4 m_\pi^2, t_+)$  from two different analyses: the recent dispersive treatment reported in~\cite{Schneider:2012ez} and  the older work~\cite{Koepp:1974da}. For the quantity $I$ we used the estimate given in~\eqref{eq:Ierror}.
It turns out that the value of the form factor at $t=0$  plays a significant role in producing stringent constraints. 
Although in principle  $f_{\omega\pi}(0)$ can be complex, we assumed that it only has a small phase which can be neglected~\cite{Schneider:2012ez}.
The upper and lower bounds given below were obtained using as input the central value  $f_{\omega\pi}(0)=2.30\gev^{-1}$. 

We checked that the bounds are quite stable with respect to the variation of the input: by varying $f_{\omega\pi}(0)$ inside the error quoted in~\eqref{eq:f0}, the upper bounds in the region of interest are changed by at most 2.5\%. Also, the uncertainty of the quantity $I$ quoted in (\ref{eq:Ierror}) affects the bounds by at most 2\%.  As the experimental errors are currently rather in the 10--20\% range, we refrain from displaying these small variations in the bounds graphically and only discuss the central results.

\begin{figure}
\centering
\includegraphics[width=\linewidth,clip]{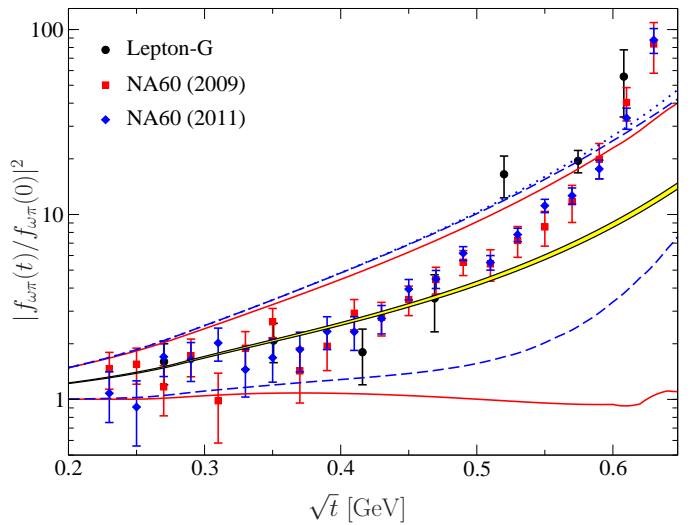}
\caption{  Upper and lower bounds compared with experimental data on $|f_{\omega\pi}(t)/f_{\omega\pi}(0)|^2$. Solid red line: bounds calculated using the expressions~\eqref{eq:rangef} with input from~\cite{Schneider:2012ez}. Dashed blue: optimal bounds calculated using~\eqref{eq:boundreal} with input from~\cite{Koepp:1974da}. The dotted blue line is the 
upper bound calculated with the same input~\cite{Koepp:1974da}, but using the nonoptimal expression~\eqref{eq:rangef}.
The data are from Lepton-G~\cite{LeptonG}, NA60~(2009)~\cite{NA60}, and NA60~(2011)~\cite{NA60new}.  The yellow band is the result of the dispersive calculation performed in~\cite{Schneider:2012ez}.
\label{fig:1}}
\end{figure}

\begin{sloppypar}
Our results are presented in Fig.~\ref{fig:1}, which shows upper and lower bounds on the modulus squared (normalized to its value at $t=0$) in the part of the elastic region accessible experimentally in $\omega\to\pi^0\mu^+\mu^-$. For the input from~\cite{Schneider:2012ez}, when the form factor is not a real analytic function, the  bounds on $|f_{\omega\pi}(t)|$ were calculated using~\eqref{eq:rangef}. For the input from~\cite{Koepp:1974da}, where rescattering effects are neglected and the form factor is real analytic,  we used the optimal bounds~\eqref{eq:boundreal} on the real part, and combined them with the knowledge of the imaginary part to obtain bounds on the modulus.  To assess the loss of optimality inherent in~\eqref{eq:rangef}, we also show the upper bound calculated with this expression for the input from~\cite{Koepp:1974da}. One can see
  that the bounds calculated  using~\eqref{eq:rangef} are very close to
the optimal bounds calculated with~\eqref{eq:boundreal} 
for the elastic energy range of interest.   For comparison, we also show the result of the dispersive calculation performed in~\cite{Schneider:2012ez}, as well as several experimental data from~\cite{LeptonG,NA60,NA60new}. 

   Figure~\ref{fig:1} shows that, although the allowed ranges for the ratio 
$|f_{\omega\pi}(t)/f_{\omega\pi}(0)|^2$ calculated with~\eqref{eq:rangef} and~\eqref{eq:boundreal}
are rather large, the upper bounds exclude some of the data points from~\cite{LeptonG,NA60,NA60new}  in the region above $0.6\gev$. Specifically, the exclusion is at the $4\sigma$ level for the last point (at $0.63\gev$) from~\cite{NA60new}, and at about the $2\sigma$ level for the other points lying above the upper bounds. On the other hand, the dispersive calculation performed in~\cite{Schneider:2012ez} is situated in the allowed range for the modulus derived here. 
Note also that in~\cite{Terschluesen:2010ik,Terschlusen:2012xw}, $f_{\omega\pi}(t)$ has been calculated based on a low-energy effective theory including explicit vector-meson degrees of freedom. We have checked that, although the representation~\cite{Terschluesen:2010ik,Terschlusen:2012xw} rises more quickly than the dispersive ones~\cite{Schneider:2012ez,Koepp:1974da} and follows the data more closely than most others, it still lies comfortably inside the bounds.
\end{sloppypar}

\section{Discussion and conclusions}\label{sec:disc}

The present study was motivated by the discrepancies noticed recently between the theoretical calculation of the $\omega\pi$ form factor by a dispersion relation and some of the data reported in~\cite{LeptonG,NA60,NA60new}. Our aim was to avoid the assumptions made on the discontinuity of the form factor above the threshold $t_+$, where the elastic unitarity~\eqref{eq:disc} is no longer valid. To this end we have resorted to the formalism of unitarity bounds. 
The central point of the formalism is the derivation of an integral condition on the modulus squared of the form factor from $t_+$ to infinity, which can be calculated using OPE and perturbative QCD in the Euclidean region for a suitable correlator,  together with unitarity and positivity of the spectral function. In the present case it was convenient to consider the tensor~\eqref{eq:correl} of two isovector currents.\footnote{It is easy to see that, if one considers the full electromagnetic current in~\eqref{eq:correl}, the QCD contribution~\eqref{eq:qcd} will be scaled by a factor greater than unity (equal to 4/3), and therefore the formalism will lead to weaker bounds. They can be improved if one includes the most important low-energy isoscalar states, approximated by narrow $\omega$ and $\phi$ resonances, in the unitarity sum,  which will be subtracted along with the two-pion contribution as in~\eqref{eq:I}. As the bounds obtained in the present work correspond to subtracting the \emph{full} isoscalar contribution, they are obviously slightly stronger than those based on the full current in such a procedure.} From the integral condition, by using techniques of analytic interpolation theory~\cite{Duren}, one can derive bounds on the form factor or its derivatives at points inside the holomorphy domain. 

In this work we have considered a modified version of the standard formalism, suitable for including the information available on the $\omega\pi$ form factor, namely the discontinuity~\eqref{eq:disc} known in the elastic region. For real analytic functions, the bounds~\eqref{eq:boundreal} are consistent with the more general results given in~\cite{Caprini:1981,Caprini:1982nr}. We have also derived bounds on the modulus of the form factor in the case when, due to rescattering effects, it is not an analytic function of real type.
This is a generalization of the formalism of unitarity bounds, considered for the first time in this paper. The framework is not specific to the $\omega\pi$ form factor and could easily be adapted to the
analysis of other form factors. The derivation 
uses the maximization of the modulus of a difference of complex numbers in the last step, which implies that strictly speaking the bounds~\eqref{eq:rangef} are not optimal. In practice, for the real analytic form factor with the discontinuity from~\cite{Koepp:1974da}, the predictions of~\eqref{eq:rangef} are almost indistinguishable from the optimal bounds~\eqref{eq:boundreal} in the energy region of interest. 

The numerical results 
show that several experimental data around $0.6\gev$ are situated above the upper bounds derived in the present paper, using two different evaluations of the discontinuity~\eqref{eq:disc} in the elastic region. 
Having in view the model-independent treatment of the region above the inelastic threshold $t_+$ adopted in our analysis, the disagreement signals possible problems with the experimental data.  
Adding to the seeming inconsistency between data on $f_{\omega\pi}(t\leq t_-)$ from $\omega\to\pi^0\ell^+\ell^-$
and data on $f_{\omega\pi}(t\geq t_+)$ from $e^+e^-\to\omega\pi^0$~\cite{SND-omegapi,CMD2-omegapi,KLOE:omegapi0gamma} (see e.g.\ the compilation in~\cite{Czerwinski:2012ry} or the continuation of the calculation~\cite{Schneider:2012ez} above $t_+$ shown in~\cite{Schneider:Diss}), our results therefore strongly support the desirability of renewed experimental efforts to measure the $\omega$ conversion decay in a more exclusive setting~\cite{WASA,CLAS}.

\begin{acknowledgement}
\begin{sloppypar}
\noindent{\bf Acknowledgements} \  We
would like to thank Stefan Leupold for useful comments and suggestions.
Financial support 
by the Ministry of Education (Contract PN 09370102/2009), by CNCS-UEFISCDI (Program Idei,  Contract 121/2011),
by the DFG (SFB/TR 16, ``Subnuclear Structure of Matter''),
and by the project ``Study of Strongly Interacting Matter'' 
(HadronPhysics3, Grant Agreement No.~283286) 
under the 7th Framework Program of the EU
is gratefully acknowledged.
\end{sloppypar}
\end{acknowledgement}

\end{document}